# SYSTEM ON PROGRAMMABLE CHIP FOR PERFOMANCE ESTIMATION OF LOOM MACHINE

Gurpreet Singh[1] , Ajay Kumar Roy[2] , Surekha K S[3] , S Pujari[4]

[1]Infosys, Chandigarh, India
*jatt_gurpreet@yahoo.com*
[2]Infosys, Hyderabad, India
*ajayroy70@yahoo.co.in*
[3]Army Institute of Technology, Dighi, Pune, India
*surekhaks@yahoo.com*
[4] SUIIT, Sambalpur University, sambalpur, Odisha, India
*sspujari@hotmail.com*

## ABSTRACT

*System on programmable chip for the performance estimation of loom machine, which calculates the efficiency and meter count for weaved cloth automatically. Also it calculates the efficiency of loom machine. Previously the same was done using manual process which was not efficient. This article is intended for loom machines which are not modern.*

## 1. INTRODUCTION

Now a days the loom machines used are very sophisticated and costly [1] [8][9][11]. So small factories running loom machine are not able to purchase those costly loom machines. So we thought of upgrading the older loom machines which are still in use in small factory like one in Kolhapur, a city in the state of Maharashtra, India. We are making it easier to calculate the performance parameters like efficiency calculation, meter calculation etc. Traditionally calculation of efficiency is a complete manual process. Here overall meters of cloth weaved by a machine in one shift will be considered. For calculations of efficiency there were various disadvantages like loss of data, error due to machine stoppage, manual error etc.

In current PC based electronic system *loom machine* there are certain advantages over traditional system. But in spite of that there are several disadvantages like PC should always be ON, complete utilization of PC is not done, and the need of skilled labors.

In the new proposed system we are using VLSI technology instead of microprocessor. This system has an optional PC interface which will be used for data storage. Since the PC interface is optional, system functioning will be independent of PC. This system has a user friendly keyboard which can be operated by any person. It is not mandatory that the operator should have prior knowledge of PC.

   

## 2. BLOCK DIAGRAM

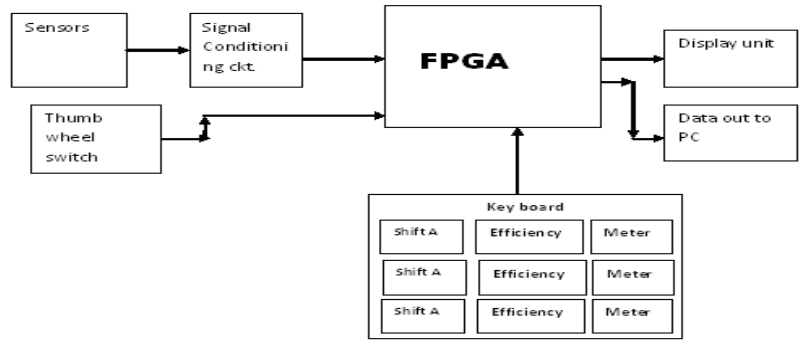

Figure no. 1

The block diagram is as shown in figure 1.

## 3. EXPLANATION OF BLOCK DIAGRAM

### 3.1. HALL-EFFECT PROXIMITY SWITCH SENSOR (HP)

Non-contact magnetic sensor proximity switch produces a digital output. The output switches between logic low (operate point) and logic high (release point) with presence and absence of a magnet as a target. The difference between the magnetic operate and release points are called the hysteresis of the device. The built-in hysteresis circuitry allows clean switching of the output even in the presence of external mechanical vibration and electrical noise [2].

### 3.2. SENSOR

The sensor as shown in figure no. 1 is connected to the rotating shaft and for that signal is generated and then it is amplified and converted to square wave using signal conditioning circuit, and it is given to FPGA.

### 3.3. THUMB WHEEL SWITCH

This will decide the pick value. We will be using thumbwheel switch (3 wheels) and it will decide the density of Horizontal threads per inch [3]. A typical thumbwheel switch is shown in figure no. 2

### 3.4. KEYBOARD

The keyboard is connected to the FPGA which contains the labels shift A, shift B and shift C, now by pressing the "efficiency button" it will display the efficiency.

### 3.5 DISPLAY UNIT

For display we are using seven segment display as shown in the modified block diagram . See figure no. 4.

### 3.6. DATA OUT PC: UART

The Universal Asynchronous Receiver / Transmitter (UART) controller is the key component of the serial communications subsystem of a computer. The UART takes bytes of data and transmits the individual bits in a sequential fashion.we are using only the serial transmitter in this project [4].

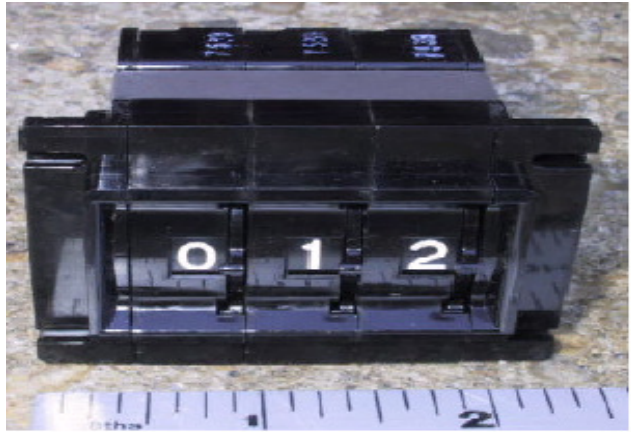
Thumb Wheel Switch

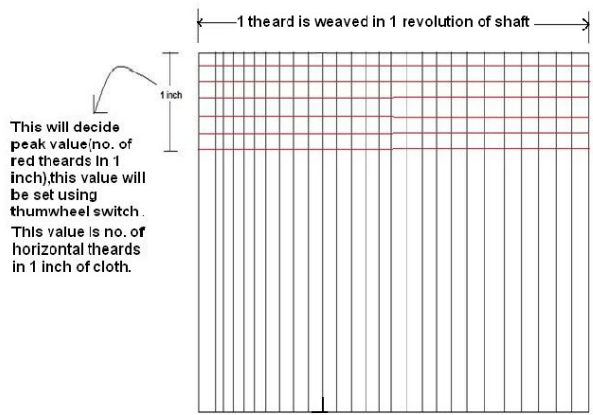

## 4. INSIDE FPGA

a) Input signal from the sensor is given to counter, the rotation counter saves the data of shift A, shift B, shift C in buffer (i.e total no. of rotations is equivalent to total no. of horizontal threads).

b) The upper binary converter is used to convert the decimal RPM value of machine shaft (which is specified) into binary value for further calculations.

c) By using the data of buffer and the binary converter value we can calculate the efficiency and meters of cloth, with the help of formula provided.

d) Now the data from the efficiency calculator and meter calculator is stored in another buffer and it will be displayed on the display unit.

### 4.1. Rotation Counter

This will count the no of rotations of the shaft

### 4.2. Binary Converter

It is used to convert the decimal RPM value of machine shaft (which is specified) into binary value for further calculation.

### 4.3. Buffer for Storing Total Rotation Data

These will store the total no of rotations of shaft (i.e. total no of Horizontal threads used in particular shift).
BUFFER VALUE IS:
999*60* SHIFT HOURS (999 is max. speed of shaft in rpm)
MAX BUFFER VALUE IS 719280

### 4.4. Efficiency Calculator

This will calculate the efficiency
*Efficiency* = [Count value/ (rpm* 60 * shift hours)]*100

### 4.5. Meter Calculator

It will calculate the total length of cloth in meter.
*Meter* = Count value / (pick value *39.37)

## 5. DESIGN APPROACH

i. In this we will be using FPGA XC3S200 FT256 kit with in built serial port [5].
ii. We will mount sensors, thumbwheel switch, and all other on a zero PCB
iii. Data will be sent to PC using a serial cable
iv. Seven segment display will be used to show current meter calculations
v. At the end of shift, just by pressing efficiency button one can get efficiency value

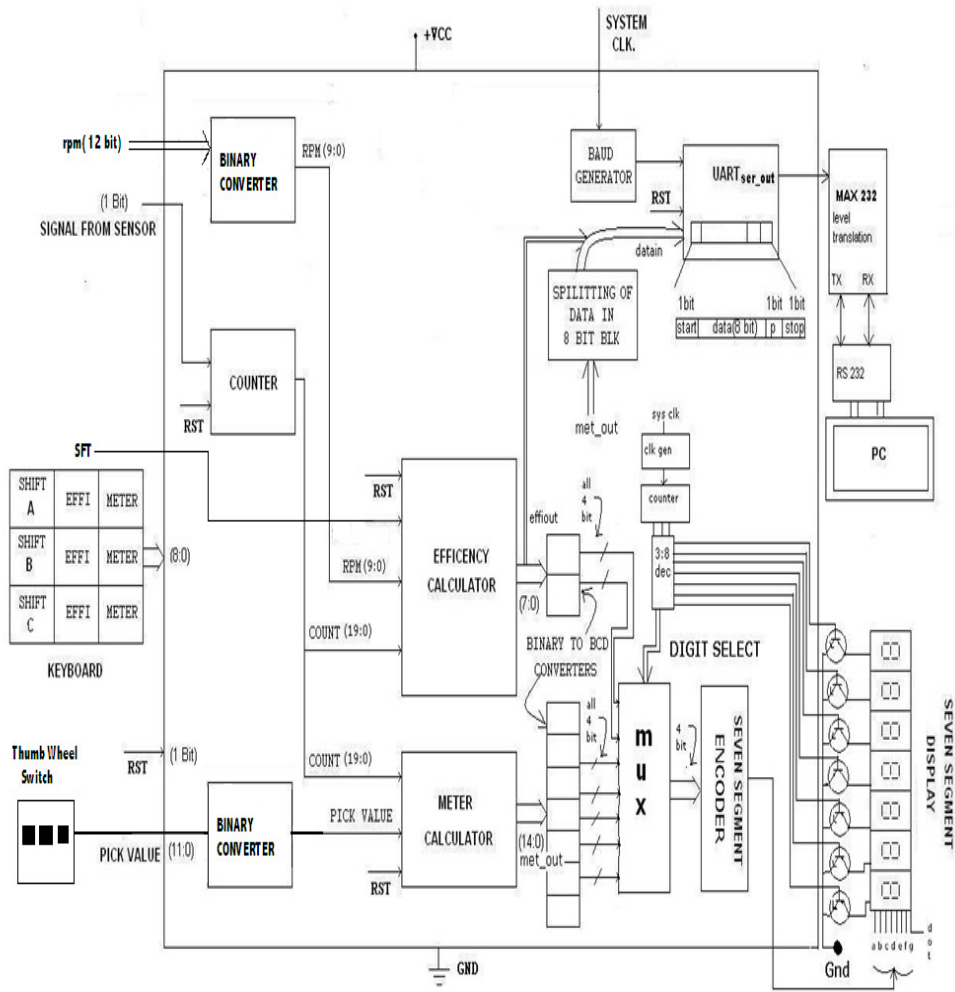

Figure No. 4. MODIFIED BLOCK DIAGRAM

## 6. SOFTWARE REQUIREMENTS

[1] VHDL or Verilog programmer & Editor
[2] Xilinx ISE
[3] Modelsim

## 7. HARDWARE REQUIREMENTS

[1] FPGA XC3S200 FT256
[2] PC with RS-232 cable.
[3] General FPGA Kit.
[4] RPM Counter Proximity Sensor
[5] Signal Conditioner
[6] Thumbwheel switch
[7] Seven Segment Led Display Unit
[8] Push Button switches

## 8. SIMULATION RESULT OF SOME OF THE MAIN MODULES

### 8.1 EFFICIENCY CALCULATOR

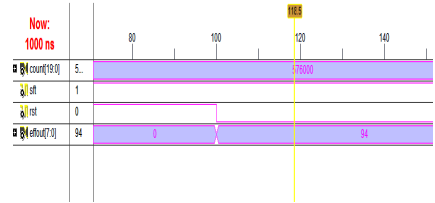

**Figure No. 5**

From figure no.5 we can see the Results of Efficiency Calculator

### 8.2 METER CALCULATOR

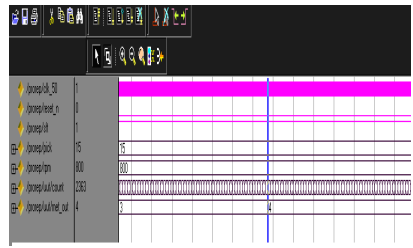

**Figure No. 6**

### 8.3 UART SIMULATION

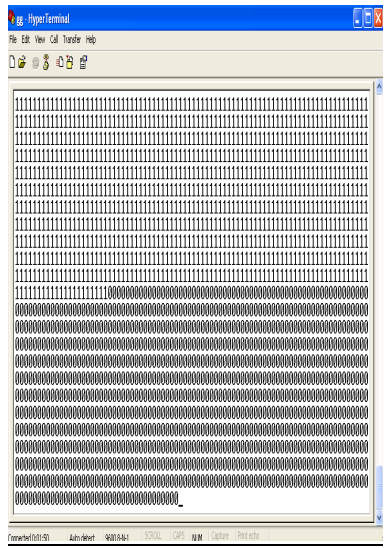

**Figure No. 7**

From transmission of figure no. 7 we can see the results of UART simulations with numeral 1 and 0 on Hyper Terminal of the PC.

## 9. FUTURE SCOPE

[1] GENERALIZED KIT: hardware can be modified itself on the kit.
[2] Data stored on PC can be made available on internet[10][12] for its remote usage
[3] Can include various other issues  related to loom machine such as:
a) Speed control of loom machine      b) Selection of horizontal threads

## 10. CONCLUSION

The system uses VLSI technology instead of microprocessor. PC is used as an optional interface and  will be used for data storage. Since the PC interface is optional, the system functioning is independent of PC. The user friendly keyboard used in the system can be  operated by any person. It is not mandatory that the operator should have prior knowledge of PC. System on programmable chip (SOPC) for performance estimation of loom machine**,** calculates the efficiency and meter count for weaved cloth automatically. Previously same was done using manual process which was not efficient. The data available on PC can be made available on internet for remote usage.